\documentstyle[12pt]{article}

\def\Journal#1#2#3#4{{#1} {\bf #2}, #3 (#4)}

\def\PLB{{\em Phys. Lett.}  B}

\def\PRL{\em Phys. Rev. Lett.}

\def\PRD{{\em Phys. Rev.} D}

\def\ra{\rightarrow}

\def\al{\alpha}

\begin{document}

\topskip 2cm 
\begin{titlepage}

\hspace*{\fill}\parbox[t]{4cm}{EDINBURGH 96/6\\ June 1996}

\vspace{2cm}

\begin{center}
{\large\bf Large Rapidity Gaps between Jets at HERA and at the Tevatron} \\
\vspace{1.5cm}
{\large Vittorio Del Duca} \\
\vspace{.5cm}
{\sl Particle Physics Theory Group,\,
Dept. of Physics and Astronomy\\ University of Edinburgh,\,
Edinburgh EH9 3JZ, Scotland, UK}\\
\vspace{1.5cm}
\vfil
\begin{abstract}
\noindent
In this talk I consider 
the formation of a rapidity gap in hadron production between two jets.

\end{abstract}

\vspace{2cm}

{\sl To appear in the Proceedings of\\ the International Workshop on\\
Deep Inelastic Scattering and Related Phenomena\\
Roma, Italy, April 1996}
\end{center}

\end{titlepage}

\baselineskip=0.8cm


\noindent
Rapidity gaps in hadron production between two jets
have been observed in photoproduction at HERA~\cite{hera} and in $p\,\bar p$
collisions at the Tevatron~\cite{tev}, where it was
found that collecting events with two or more 
jets and ranking them according to the rapidity interval $\Delta\eta$ between
the two jets with the largest transverse energy (leading $E_T$ jets), roughly 
1~\% of the events showed a gap in particle production 
between the leading $E_T$ jets. The ZEUS Collaboration~\cite{hera} 
has observed the same phenomenon in photoproduction, but at
a higher rate (roughly 7~\% of the events show a gap).

The original theoretical motivation for examining two-jet production 
with a rapidity gap was to find a clear signal in $p\, p$ collisions at
the SSC/LHC Colliders for the production of a
Higgs boson heavy enough to decay to a pair of $W$ or $Z$ bosons.
The leading Higgs-boson production mode is via gluon-gluon fusion,
however it will be very difficult to distinguish the signal from the
$t\,\bar t$ and $W\,W$ QCD backgrounds. The $W\,W\ra H$ production rate
is typically smaller, but it shows a distinct pattern of soft-hadron
and minijet production in the central-rapidity region, since it is
characterized at the parton level by color-singlet exchange in the cross
channel \cite{dkt,bj}.

Accordingly, as a preliminary investigation it has been proposed \cite{bj}
the study of color-singlet exchange between
two partons, which can be carried over at the existing colliders. At the
parton level the formation of a rapidity gap may be realized via the exchange
in the cross channel of either an electroweak boson or of two gluons in a 
color-singlet configuration, which occurs at ${\cal O}(\al_s^4)$. 
However, the gap-production rate due to 
electroweak-boson exchange is in this case rather small and can be neglected 
\cite{che}. The exchange of a gluon, which is an ${\cal O}(\al_s^2)$ 
process, constitutes the background. The probability for it
to yield a rapidity gap is suppressed
by a Sudakov form factor which makes it fall off exponentially as the 
rapidity gap between the partons widens, because the exchanged gluon is
likely to radiate off more gluons in the gap \cite{ddt}. 
Therefore a rapidity gap between
two partons is an indication for a strongly interacting color-singlet
exchange. Bjorken \cite{bj} estimated that $\hat f_{gap}\equiv 
{\hat\sigma_{sing}/\hat\sigma_{oct}}\sim 0.1$.

However, to produce a gap at the parton level is not sufficient because 
in the hadronization the gap is usually filled by the hadrons emitted in
the spectator-parton interactions of the underlying event. 
Bjorken \cite{bj} estimated the rapidity-gap survival probability, $<|S^2|>$,
to be about 5-10\%. In a first approximation we can then assume that the 
fraction of two-jet events with a gap in soft-hadron production is given by
$f_{gap}\simeq <|S^2|> \hat f_{gap}$. Thus gaps were predicted at about 
the 1\% level~\cite{bj}, in qualitative agreement with the Tevatron experiments
\cite{tev}. In photoproduction, though, the ZEUS Collaboration~\cite{hera}
observes gaps at about the 7\% level. There are two possible explanations
for the difference: at the hadron level $<|S^2|>$ is expected to increase 
as the c.m. energy $\sqrt{s}$ decreases \cite{bj,glm}, and 
$\sqrt{s_{\gamma\, p}}$ at HERA is about an order of magnitude smaller than 
$\sqrt{s_{\bar p\, p}}$ at the Tevatron; at the parton level there are 
contributions to the gap from a resolved photon, for which two-gluon 
exchange is just like in $\bar p\, p$ collisions, and from a direct photon, 
which splits into a $q\,\bar q$ pair which then exchanges
two gluons with a parton within the proton. Both contributions are
${\cal O}(\al_s^4)$, and mix together. The direct-photon component, though,
has $<|S^2|>=1$ since the photon is point-like. The combination of these 
two effects may account for the difference between the gap rate at HERA and 
at the Tevatron. Any quantitative analysis, though, is not possible
until ${\cal O}(\al_s^4)$ calculations are carried out in detail.

\section*{Acknowledgments}
I thank PPARC and the Travel and Research Committee of the
University of Edinburgh for the support, and the organizers of DIS96
for the hospitality.


\end{document}